\documentclass[english,aps,prper,reprint,showpacs,titlepage,longbibliography]{revtex4-2}   

\usepackage[T1]{fontenc}	
\usepackage{geometry}
\usepackage{graphicx}
\usepackage{times}
\usepackage{hyperref}  
\hypersetup{colorlinks=true,urlcolor=blue,citecolor=blue,linkcolor=blue}
\usepackage{array}
\usepackage{enumerate}
\usepackage{enumitem}  
\usepackage{amsmath}
\usepackage{amssymb}
\usepackage{tikz}
\usepackage{graphicx}
\usepackage{multirow}
\usepackage{tcolorbox}
\usepackage{ragged2e}
\usepackage{tcolorbox}
\usepackage{float}
\usepackage[utf8]{inputenc}

\usepackage[normalem]{ulem}

\begin{document}
\begin{titlepage}

\title{Cognitive Load and Situational Interest in Physics Laboratories: A Comparative Study Across Three Instructional Modalities}

 \author{Razan Hamed}
 \affiliation{Department of Physics and Astronomy, Purdue University, 525 Northwestern Ave, West Lafayette, IN, 47907, U.S.A.} 
 
 \author{N. Sanjay Rebello}
\affiliation{Department of Physics and Astronomy, Purdue University, West Lafayette, IN, 47907, U.S.A.\\
 Department of Curriculum and Instruction, Purdue University, West Lafayette, IN, 47907 U.S.A.}

\keywords{}

\begin{abstract}

Understanding how an instructional approach shapes students cognitive resources and engagement is central to improving undergraduate physics education especially for novice learners. This study examines how three instructional modalities (Inquiry-based, Design-based, and Game-based learning) affect cognitive load and situational interest in physics laboratories for non-STEM majors. Guided by the revised Cognitive Load Theory framework, two experiments were conducted across two physics domains: mechanics and electrical circuits. In each experiment, students completed three laboratory sessions, one in each instructional modality, followed by surveys measuring cognitive load and situational interest. One-way ANOVA analyses revealed significant differences across the three modalities in both experiments. Game-based laboratories consistently yielded the lowest cognitive load and the highest situational interest, while inquiry-based and design-based labs imposed higher cognitive demands, with their relative effects varying by domain. Overall, situational interest exhibited an inverse relationship with cognitive load, suggesting that reduced cognitive demands support greater engagement. These findings emphasize the value of strategically selecting and combining instructional modalities to balance cognitive load and foster meaningful engagement in physics laboratories for novice learners.

\clearpage
\end{abstract}

\maketitle
\end{titlepage}

\section{Introduction}

In the evolving landscape of physics education, instructional development plays a pivotal role in shaping how physics learners process, internalize, and engage with complex disciplinary content. The integration of cognitive load theory and situational interest frameworks into physics instruction for non-STEM majors provides a dual lens for understanding not only the mental effort required for learning but also the affective responses that sustain students motivation in physics. 

In this study, we examine how three different instructional modalities (Design-based, Inquiry-based, and Game-based) affect novice physics learners' multidimensional cognitive load and situational interest. Specifically, we aim to answer the following research questions:

\begin{enumerate}[]
\item[\bf RQ 1:] {How do the three instructional modalities (Design-based, Inquiry-based, and Game-based) impact students’ cognitive load and situational interest?}
\item[\bf RQ 2:] {How does cognitive load and situational interest differ across the three instructional modalities within the contexts of mechanics and circuits?}
\end{enumerate}

\section{background}
The realm of physics education has been explored from many perspectives and frameworks. This study is centered on one such framework: Cognitive Load Theory (CLT). The purpose of Cognitive Load Theory is to investigate the working memory load and how it affects learning of new information \cite{sweller1988cognitive}. In addition to cognitive load, context-based interest in the instructional material have also been shown to impact student learning significantly \cite{kapp2012gamification}. Therefore, this study focuses on analyzing both the cognitive load and  situational interest of non-STEM physics learners in the context of various instructional methods commonly used in physics. Specifically, we compare three instructional approaches to learning physics: Inquiry-based, Design-Based, and Game-Based Physics Learning. 

Previous studies have shown that Cognitive Load Theory (CLT) is a useful and influential framework for studying how students learn and solve problems \cite{sweller2022role}. CLT focuses on how limitations of working memory capacity and human cognitive architecture affect learning, particularly for novice learners \cite{sweller2011cognitive}. In its initial version CLT proposed that learning outcomes are influenced by three distinct types of mental load: Intrinsic Cognitive Load (ICL), driven by the inherent complexity of the material; Extraneous Cognitive Load (ECL), shaped by the instructional design; and Germane Cognitive Load (GCL) reflecting the learner’s mental effort devoted to constructing meaningful understanding \cite{sweller1988cognitive}. 

The intrinsic cognitive load pertains to the challenges inherent to the difficulty of the learning material and is determined by the number of interacting elements that the learner has to keep in mind \cite{sweller1988cognitive}. Germane cognitive load is associated with the level of focus and effort exerted by learners and it is influenced by the type of instructional method and the motivation of the learner \cite{sweller2003evolution}. Extraneous Cognitive Load is related to unnecessary and irrelevant information that the learners must interact with due to the chosen method of instruction. Extraneous load can negatively impact the learning outcome of the students as it is not directly relevant to the learning goals \cite{sweller1988cognitive}.

While cognitive load theory has long been understood in terms of intrinsic, germane, and extraneous cognitive load, recent theoretical developments have led to revisions reshaping the focus of CLT and categorization of cognitive load \cite{kalyuga2025rethinking}. CLT initially was comprised of only two types of cognitive load (Intrinsic and Extraneous) and later modifications included the introduction of germane cognitive load to explain why certain learning activities can improve learning outcomes despite increasing the mental effort involved in learning\cite{kalyuga2011cognitive}. Namely, the Germane load was proposed to capture cognitive processes that intentionally support schema construction beyond the basic task demands represented by the intrinsic load. Over time, however, substantial conceptual overlap emerged between intrinsic and germane load, as both were defined in terms of essential processes for learning \cite{kalyuga2011cognitive, debue2014does}. This overlap created theoretical ambiguity and measurement difficulties, ultimately leading to a recent re-conceptualization of CLT that abandoned the germane load in favor of a clearer dual model consisting only of intrinsic and extraneous cognitive load, with intrinsic load broadly defined by the learner's interest and essential cognitive activities required to achieve the learning goals \cite{kalyuga2025rethinking}. Therefore, to align with the recent revisions of CLT, this study focuses exclusively on the intrinsic and extraneous cognitive load along with examining the situational interest of physics learners. 

Since intrinsic and extraneous cognitive load are crucial to the learning process, effective instruction seeks to reduce ECL and manage ICL in order to maximize learning and minimize irrelevant cognitive demands \cite{sweller1988cognitive}. Therefore, it is important to consider how instructional methods in physics affect those two types of cognitive load. Research has found that elements aimed at increasing motivation or inducing positive emotions are considered part of intrinsic cognitive load if they align with the learner's needs and the activity's goals. Conversely, the same elements become extraneous for a learner who is already motivated or emotionally engaged \cite{plass2019four, hawthorne2025relationship}.

Complementary to cognitive load is the concept of situational interest; while cognitive load pertains to the working memory capacity, situational interest is concerned with the learners’ immediate affective response and attention directed at the task, thus it can highly impact learning outcomes\cite{linnenbrink2010measuring}. Situational Interest (context-specific interest) differs from individual interest which is the intrinsic desire to engage in a particular activity regardless of the context. \cite{hidi2004interest, renninger2011revisiting, schiefele1991interest} Since situational interest enhances value attribution to the learning experience, it is expected that the situational interest generated in learners will develop over time into individual interest in the learning task \cite{rotgans2011situational}. Therefore, situational interest is critical not only for sustaining focus during instruction but also for shaping long-term attitudes toward learning in STEM disciplines \cite{linnenbrink2010measuring}.

This study compares the cognitive load and situational interest of physics learners across three different types of instructional methods: inquiry-based learning, design-based learning, and game-based learning. While each of these modalities have been studied in the literature, they have seldom been compared with each other with respect to their cognitive load and situational interest. These three instructional modalities have also been defined in different ways and no unique definition of each exists in the literature. Therefore, for the purposes of this study, we provide an operational definition of each modality.

Inquiry-based learning entails self-directed exploration and involves students constructing their own mental schema of the learning material \cite{kirschner2006unguided}. Inquiry-based learning activities have been shown to place a substantial cognitive burden on learners due to their dependence on open-ended problem-solving and self-directed investigation but could potentially be useful tools for exercising autonomy in one's own learning \cite{sweller1988cognitive, kirschner2006unguided}. 

Design-based learning promotes learning through problem-solving including the iterative processes of designing, testing, and refining solutions in design activities \cite{kirschner2006unguided}. Due to the ill-structured nature of design problems, design-based learning can impose high cognitive demands on novice learners but could allow for hands-on exploration and engagement with the learning material \cite{jonassen2010learning}. 

Game-based learning utilizes educational games to help students engage and interact with the material in a fun way without compromising on problem-solving and critical thinking \cite{kapp2012gamification}. Game-based learning has been shown to positively lead to flow and reduce the cognitive burden due to its interactive and engaging nature and immediate feedback on the condition that the game-based activity is properly designed with the optimal level of challenge, clarity, and guidance \cite{chang2017development, perttula2017flow}. Therefore, effective educational games aspire to include elements that result in emotional or physical engagement that, in turn, can result in increased cognitive engagement and personal interest \cite {plass2015foundations, plass2020handbook} which are usually enhanced by productive failure and interactivity \cite{kapur2009productive, homer2014moved}.

In this study, we conduct two experiments, each concerning a distinct domain of physics: mechanics and electrical circuits. As the reader will see in the methods section, the two experiments share a common structure and differ only in the science content to be learned. The decision to include both topics in this study is largely due to the inherent nature of learning mechanics and that of learning circuits. A Syntheses of the literature on how students learn mechanics as opposed to learning circuits shows a distinct difference in the conceptual resources students tend to use when reasoning about mechanics versus circuits\cite{robertson2023identifying}. 

Research in learning of concepts in mechanics, which is a topic that lends itself to physical experiences that learners might encounter in everyday life, shows that students frequently draw on embodied everyday experiences to make sense of forces and kinematics. For example, one study conducted an experiment in which students used their own bodies to generate and regulate force in order to feel the relationship between net force and acceleration \cite{coletta2019feeling}. Another study showed how students learned mechanics by utilizing their bodies as local positioning systems where students analyze their own movements rather than an external system in the laboratory \cite{dale2020step}. Furthermore, a third study showed that students were able to map graphs onto the motion of their body relating graphs to certain bodily movements \cite {vieyra2024making}. Similarly, a study done on high school students examined how dance can serve as a source for learning physical concepts like equilibrium, center of gravity, and weight distribution \cite{solomon2022embodied}. 

In contrast, studies on learning circuits, a topic that most learners do not have direct interactions with in their day-to-day experiences, suggest that concepts like voltage, current, and power are less directly tied to immediate bodily experiences thus require different cognitive resources. For example, a study has shown that students repeatedly use connections between voltage and current or current and resistance in order to understand the terms of voltage, current, and resistance \cite{bauman2024student}. Affirming the notion that learning circuits requires distinct conceptual resources, another study has found that circuit simulations were more effective than real-life equipment for building students conceptual understanding about circuits \cite{hazelton2013assessing}. Other studies have found that circuits are better understood through the use of analogies and multiple representations to connect abstract relationships with concrete examples \cite{tomkelski2023physics, chiu2005promoting}. 

Given the different nature of learning mechanics versus circuits, this study runs two distinct experiments, one in each domain. Within each experiment, the goal is to explore how the three learning approaches (Inquiry, Design, and Game) affect the cognitive load and situational interest of novice physics learners. Although each instructional approach has been studied individually in the literature, no studies have been found in PER (Physics Education Research) literature to compare all three approaches alongside each other for the same audience, and especially for non-STEM students learning physics. Therefore, this study aims at exploring the three instructional approaches side by side in order to improve physics instruction and learning for novice learners enrolled in non-STEM majors. 

\section{Methods} 
\subsection{Context}

The participants in this study were sophomore and junior undergraduate college students enrolled in an introductory physical science course for future elementary teachers at a large U.S. Midwestern land grant university. This study involved two distinct experiments, one on the topic of Mechanics, which tends to be based on concrete and embodied experiences, and the second done on the topic of Circuits, which involves more representational abstract reasoning. 

    \begin{figure}[h!]
        \centering
        \includegraphics[width=0.5\textwidth]{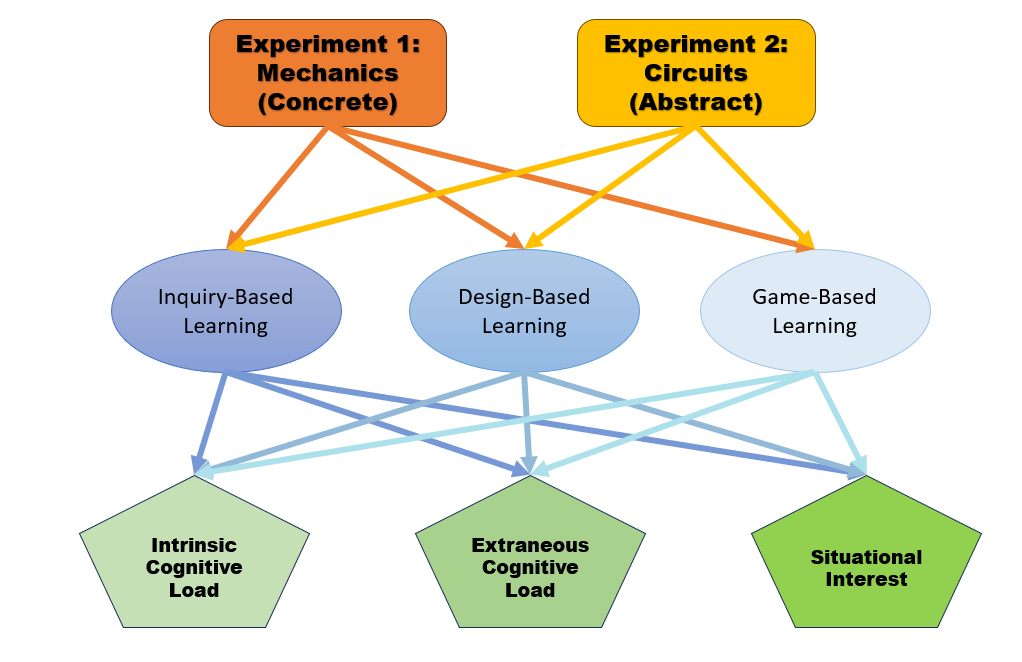}
        \caption{The flow of experiments 1 and 2}
    \end{figure}

As shown in figure 1, each experiment consisted of three laboratory sessions each corresponding to one of the three instructional modalities (Design-based, Game-based, and Inquiry-based). The same population of students completed all three laboratory sessions within each experiment since the goal is to track the changes in their cognitive load (CL) and situational interest (SI) across the three conditions. Given that both CL and SI are subjective experiences, it was important that the same students are compared across the three conditions rather than being split into three different test groups for each condition. Due to some absences in some of the laboratory sessions, the total number of students completing all three labs in each experiment varied slightly with a total of N = 54 students completing all three labs within the first topic (Mechanics) and N = 47 completing all three labs within the second topic (Circuits). 

 At the end of each lab - within both experiments - the students completed the cognitive load survey and the situational interest survey which are designed to measure cognitive load and situational interest. Both surveys have been extensively used and validated in the literature \cite{klepsch2017development, linnenbrink2010measuring}. The course was structured such that all three lab modalities were already incorporated within each of the two instructional units thus it should be noted here that the three lab types within each unit focused on similar conceptual themes, although the specific content differed so as to minimize redundancy-related effects.

\subsection{Cognitive Load Survey}

The cognitive load survey developed and validated by \cite{klepsch2017development} includes five items corresponding to the two types of cognitive load (ICL and ECL) based on the revised theory of cognitive load. Items in each type were rated on a 7-point Likert scale ranging from Strongly Disagree (1) to Strongly Agree (7). The five items were organized as follows: 

\begin{itemize}
    \item Intrinsic Cognitive Load (ICL) was assessed with two items (e.g., "While working on this lab, I had to keep many things in mind at the same time”).
    \item Extraneous Cognitive Load (ECL) with three items (e.g., “It was difficult to find the important information due to the way it was presented”).
\end{itemize}
 
\subsection{Situational Interest Scale}
Situational interest was assessed using a modified version of the Situational Interest Scale developed by \cite{linnenbrink2010measuring}. The scale includes three sub-scales:
\begin{itemize}
    \item Triggered Situational Interest (e.g., “The material presented grabbed my attention”)
    \item Maintained Situational Interest (e.g., “I enjoyed learning the material”)
    \item Maintained Situational Interest (e.g., “I think the material is important and useful”).
\end{itemize}

Each sub scale contains four items rated on a 5-point Likert scale from 1 (Not True at all) to 5 (Very True) adding up to 12 total items in the original survey. The scale used in this study was adapted from the original survey by selecting and reducing the number of items to a single item from each sub scale, resulting in a total of three items aligned with the objectives of the study. These items were selected based on their conceptual overlap with the intrinsic cognitive load, which is measured by the other survey instrument used in this study. This approach was intended to create coherence and facilitate a more direct relationship between the results obtained from the two surveys.

\subsection{Experiment 1: Three Instructional Modalities within Mechanics}

\subsubsection{Inquiry-Based Lab} 
The mechanics inquiry lab focused on energy transformations and the law of conservation of energy through a sequence of observational and simulation-based activities. Students first completed a hands-on experiment in which they released a ball from rest on an incline and qualitatively analyzed the motion and identified how kinetic and gravitational potential energy changed over time. They predicted at what points the kinetic and potential energies would be the highest. Based on their analysis of both the kinetic and potential energies, they made predictions about the total energy of the ball as it moves down the incline. 

The students were then introduced to an interactive Phet simulation \cite{energysimulation} to explore energy conservation in more controlled and idealized conditions as shown in figure 2. The simulation presented them with a skate-park model, where they examined the relationship between kinetic, potential, thermal, and total energy with and without the presence of friction. By observing energy bar graphs and motion at different positions along the track, students analyzed how energy is transferred between different forms while total energy remains conserved. 

Within the same simulation, students also had the opportunity to make changes to the skaters mass and the track shape to observe the generality of energy conservation despite those changes. Overall, the students were able to build coherent conceptual understanding of energy conservation using a combination of real-life and simulated experiments.

    \begin{figure}[h!]
        \centering
        \includegraphics[width=0.5\textwidth]{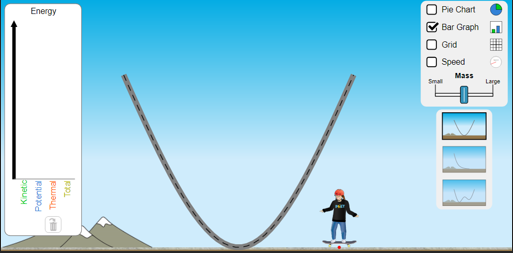}
        \caption{Inquiry-Based Lab in Mechanics}
    \end{figure}
    
\subsubsection{Design-Based Lab}
The mechanics design lab focused on applications of the conservation of energy law through designing a roller coaster that is both fun and safe as shown in figure 3. Students first engaged in an engineering design task in which they identified the design problem, established criteria and constraints related to safety and performance, and constructed a physical roller coaster prototype using foam pipes representing the tracks and a ping pong ball for the cart. The prototype was tested by releasing the ball from an elevated position on the foam pipe and observing its motion along the track, including its ability to remain on the track and safely decelerate at the end. Key physical quantities such as track height, ball mass, average speed, and travel time were measured and used to calculate kinetic energy, gravitational potential energy, and total energy at different points along the track. Comparisons of total energy values at different times allowed students to evaluate whether total energy was conserved in the system they designed and whether there was any energy loss due to friction and air resistance.

In the second part of the lab, students used a simulation of a roller coaster\cite{rollercoaster} to test their designs in a more ideal environment and validate their measurements and calculations as shown in figure 4. They explored how various elements in their design influenced total energy, speed, and acceleration. By varying parameters such as initial drop height, hill height, and cart mass while holding other factors constant, students were able to identify relationships among these variables and their effects on maximum speed and acceleration. The overall goal of the design was to create conditions that maximize fun and excitement on the ride while remaining within specified safety limits. The physical and the simulated designs were later synthesized into a final roller coaster design proposal supported by physics principles and empirical evidence. 
    \begin{figure}[h!]
        \centering
        \includegraphics[width=0.5\textwidth]{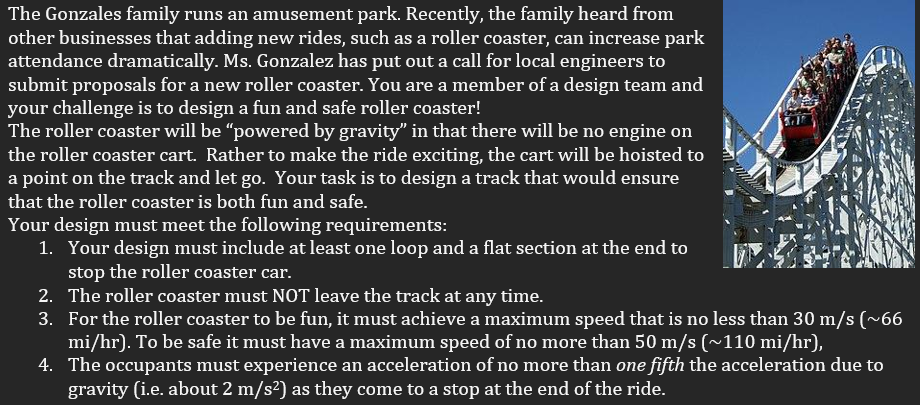}
        \caption{Design-Based Lab Prompt in Mechanics}
    \end{figure} 
   \begin{figure}[h!]
        \centering
        \includegraphics[width=0.5\textwidth]{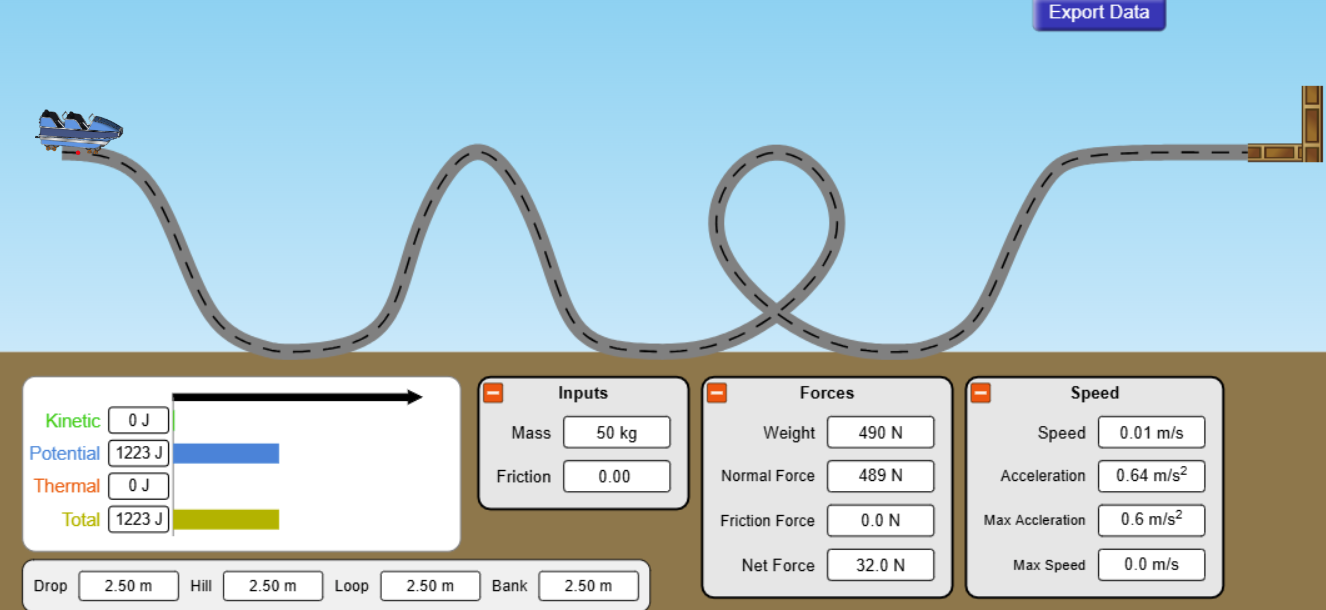}
        \caption{Design-Based Lab Simulation in Mechanics}
    \end{figure}

\subsubsection{Game-Based Lab} 
The mechanics game lab implemented an educational board game designed by the author to engage students in applying fundamental mechanics concepts related to energy and motion through structured game play. The game consisted of a physical board (Figure 5), rules sheet, calculation worksheets, player tokens, and a deck of cards containing problem-based (Figure 6) and chance-based prompts (Figure 7). Participants played in groups of 3-4 students, with each player taking turns drawing a card from the pile and responding to the scenario presented on the card. Most cards required players to calculate physical quantities such as kinetic energy, gravitational potential energy, or total energy using energy conservation equations. A smaller subset of cards introduced chance-based elements such as lucky prompts, unlucky prompts, and no action cards. All cards were formulated within light-hearted, laughter-inducing contexts and some cards incorporated elements of surprise to enhance the overall enjoyment of the game. Based on each card prompt, players advanced or retreated their tokens along the board, with the goal of reaching the finish block first. Since the prompts were presented to the players in a random order, advancing in the game was solely based on the type of calculation or chance they receive on their turn given that they perform the calculation correctly. 

Students played iteratively until one participant reached the finish block of the board, at which point the game ended. Throughout this game, players were required to document their calculations and reasoning on a provided worksheet, and group members monitored each other's work to ensure correctness of calculations and adherence to the game rules. Taking everything into account, this board game was designed to reinforce conceptual and numerical understanding of energy conservation and motion by embedding quantitative problem solving within a competitive and interactive environment. In order to engage students in a structured yet flexible way, the game integrated calculation, peer verification, and immediate feedback which are essential elements of educational game design \cite{kalmpourtzis2018educational}. 

    \begin{figure}[h!]
        \centering
        \includegraphics[width=0.5\textwidth]{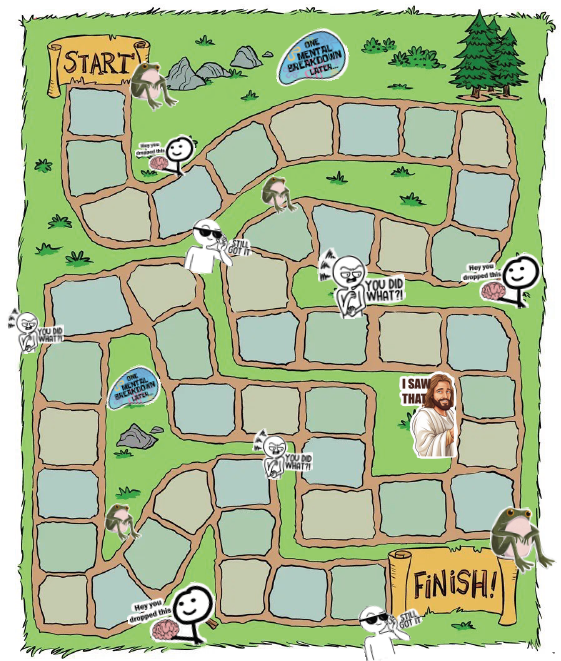}
        \caption{Game-Based Mechanics and Circuits Lab  Game Board}
    \end{figure}

    \begin{figure}[h!]
        \centering
        \includegraphics[width=0.5\textwidth]{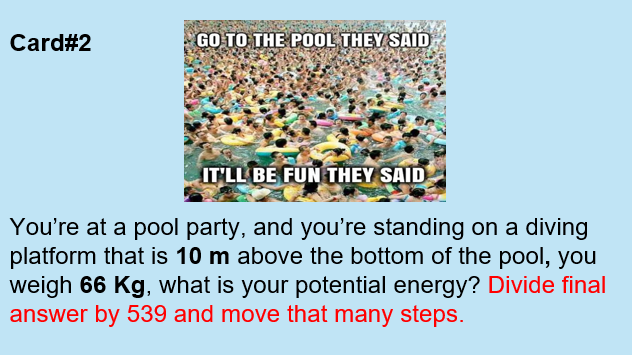}
        \caption{Game-Based Mechanics Lab Calculation Card}
    \end{figure}
    
  \begin{figure}[h!]
        \centering
        \includegraphics[width=0.5\textwidth]{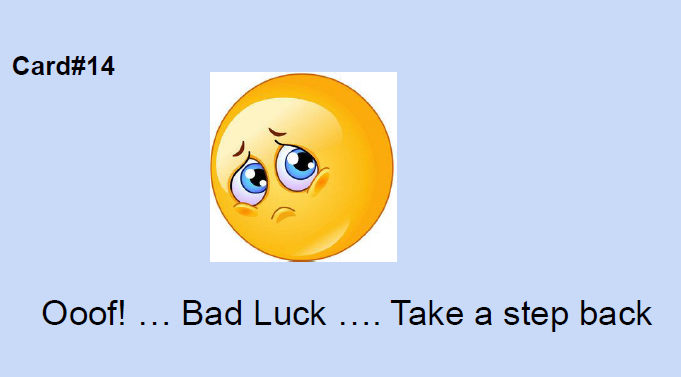}
        \caption{Game-Based Mechanics Lab Chance Card}
    \end{figure}
        
\subsection{Experiment 2: Three Instructional Modalities within Circuits}
\subsubsection{Inquiry-Based Lab}
The circuits inquiry lab introduced students to the behavior of current in combination circuits through hands-on and simulation based activities. The students had prior exposure to series and parallel circuits, therefore, in this lab they analyzed circuits composed of resistors arranged in mixed series–parallel configurations. For each circuit, they reduced the number of resistors to an equivalent resistance by learning to combine resistors correctly and calculating their total resistance using series and parallel rules. 

The students then applied Ohm’s law to determine the total current in the circuit along with individual currents and voltage values across each resistor and finally calculate the battery's power. This lab focused on reinforcing fundamental relationships between current, voltage, resistance, and power in mixed series-parallel circuits. After completing analytical and numerical calculations, students tested their results using the PhET Circuit Construction Kit \cite{circuitssimulation} as shown in figure 8. They reconstructed their circuits within the simulation environment and compared the current, voltage, and power values to the values they arrived at earlier. 

Additionally, this lab introduced students to the concept of bulb brightness through a hands-on activity (Figure 9) in which they built a circuit with multiple bulbs connected in a mixed configuration and removed certain wires and bulbs from the circuit to discover how the brightness of each light bulb had changed. Through these activities, students were able to develop a conceptual and numerical understanding of current, voltage, power, and brightness in a mixed configuration circuit. 

     \begin{figure}[h!]
        \centering
        \includegraphics[width=0.5\textwidth]{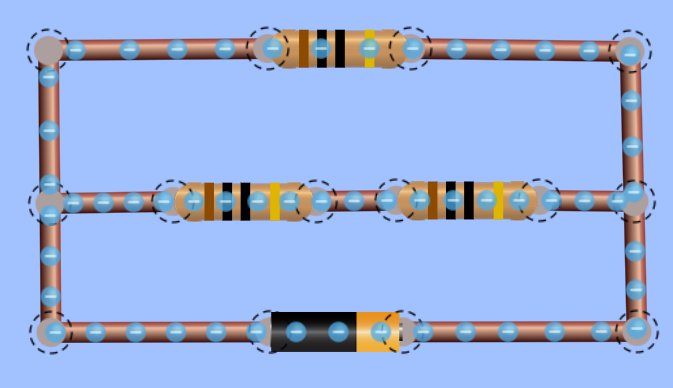}
        \caption{Inquiry-Based Lab in Circuits - Simulation}
    \end{figure}

     \begin{figure}[h!]
        \centering
        \includegraphics[width=0.5\textwidth]{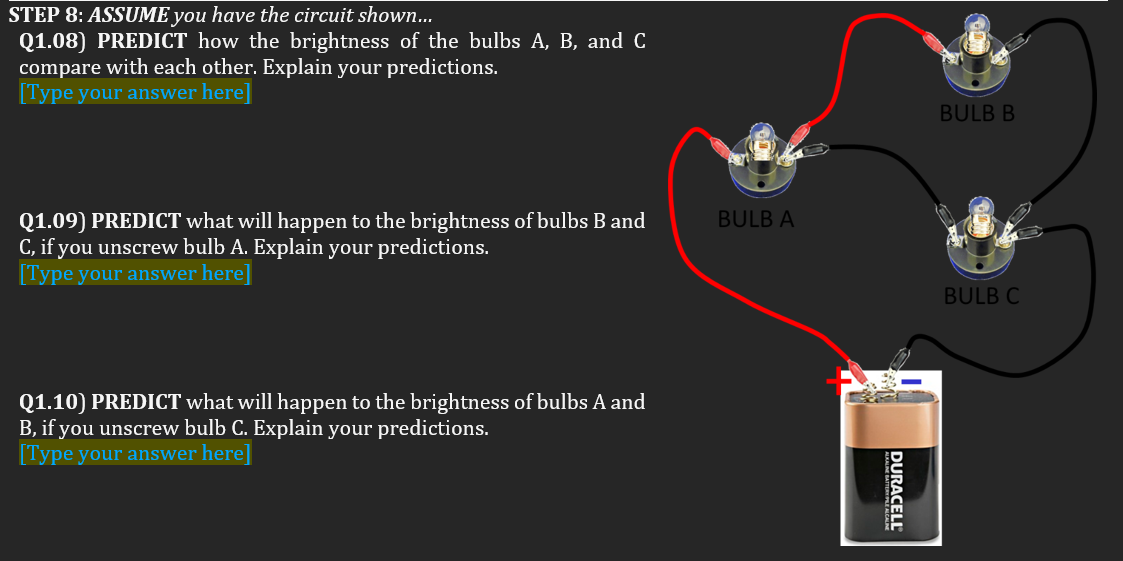}
        \caption{Inquiry-Based Lab in Circuits - Hands on}
    \end{figure}
    
\subsubsection{Design-Based Lab}
The circuits design lab implemented a practical design project focused on developing and refining an electrical alarm system with the purpose of lighting a light bulb when a locker door was open (Figure 10). This design comes as a solution to the hypothetical problem that students were presented with earlier in which their items got stolen when their lockers were left unintentionally open. Thus students were tasked with designing a functional circuit that solves the open locker problem by setting off an alarm/light that alerts students to shut their lockers. The circuit design had explicit criteria related to safety and cost efficiency while being constrained to a limited set of basic electrical components. To start the design process, students were asked to sketch prototypes of their circuits then use wires, bulbs, resistors, and a battery to build the circuits in real life. To represent the locker, students brought shoe boxes from home and arranged their circuits within the boxes according to the specified criteria. A great emphasis was placed on selecting an appropriate circuit configuration and ensuring that the light bulb only turns on when the locker's door is open.

Following their initial design, students simulated their proposed circuits using a simulation \cite{circuitssimulation} to test the functionality of their designs and verify that they met all the required criteria. Students then compared the simulated circuit to their physical prototype to identify similarities, differences, and areas for improvement. Based on the simulation results, students refined their designs to reduce electrical current and extend battery life, thereby lowering running costs while maintaining required functionality. Final designs were documented through photographs and written explanations that explicitly linked design decisions to underlying physics principles, including circuit type, current flow, and energy usage. Overall, this lab integrated hands-on and simulated design to help students apply their understanding of electric circuits to a real-life problem like open lockers.

   \begin{figure}[h!]
        \centering
        \includegraphics[width=0.5\textwidth]{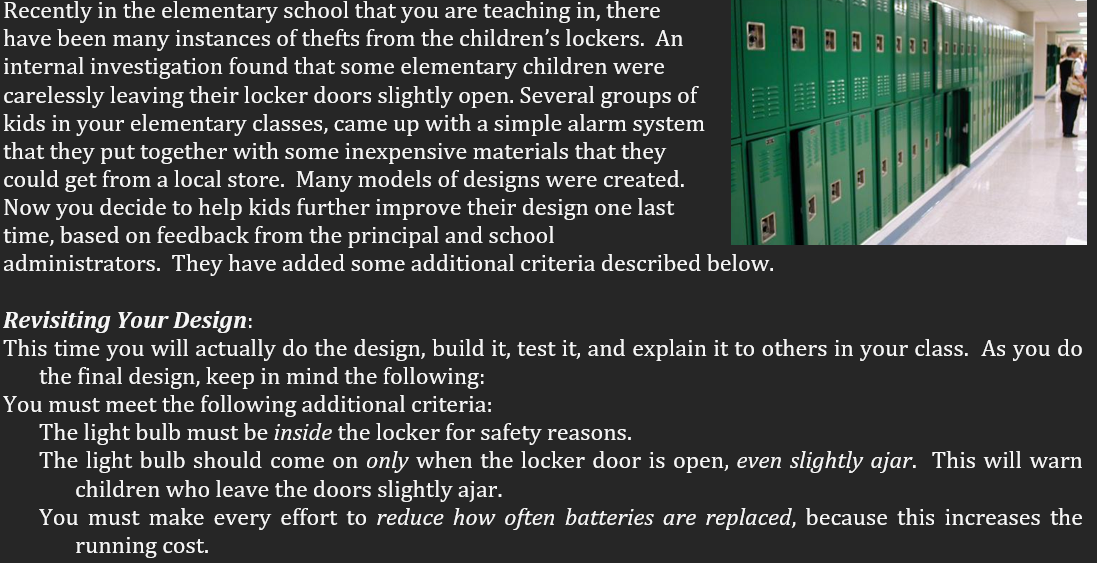}
        \caption{Design-Based Lab in Circuits}
    \end{figure}

\subsubsection{Game-Based Lab}
The circuits game lab was structured in a similar manner to the mechanics game lab. An educational board game was designed by the author to help students reinforce and apply reasoning about basic electrical circuit concepts. Just like the mechanics board game, this game was played in small groups of 3-4 students, with each participant represented by a physical token placed at the start of a track on the game board. During each round, players drew a card from a shared deck and followed the instructions provided. Most cards (Figure 11) presented short, context-based light-hearted scenarios requiring players to apply foundational circuit principles, such as Ohm’s law and power equations. They were prompted to calculate quantities like current, voltage, resistance, and power. Based on the numerical result of their calculations or, in some cases, a conceptual judgment about circuit configurations, players advanced their token a specified number of spaces. A subset of cards introduced non-calculation-based elements (Figure 12), such as lucky cards, unlucky cards, and no action cards to maintain game flow and add an element of thrill and surprise \cite{kalmpourtzis2018educational}.

Students played sequentially until one participant reached the end of the board and is declared the winner. No perks were offered to students based on their winning status to avoid taking away from the educational purpose of the game. Throughout this game, players were required to record their calculations and reasoning on a separate handout, while their peers monitored their work to ensure correctness and compliance with the rules. This game was intentionally structured so that progress was directly tied to successful application of circuit concepts with a small element of luck to make the game more engaging and more similar in nature to casual entertainment-centered games. 

 \begin{figure}[h!]
        \centering
        \includegraphics[width=0.5\textwidth]{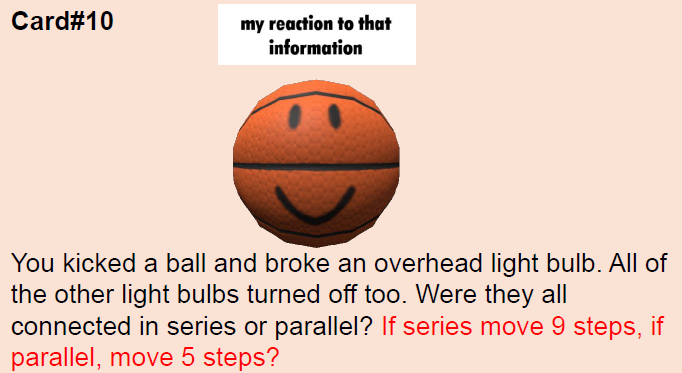}
        \caption{Game-Based Circuits Lab Calculation Card}
    \end{figure}

 \begin{figure}[h!]
        \centering
        \includegraphics[width=0.5\textwidth]{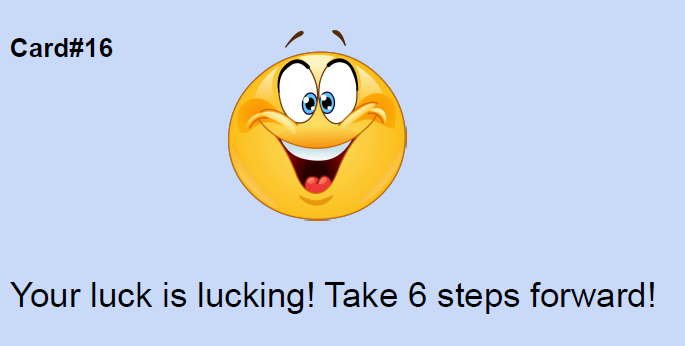}
        \caption{Game-Based Circuits Lab Chance Card}
    \end{figure}

\section{Data Collection}
At the end of each lab session, students completed both the cognitive load Survey and the situational interest survey \cite{klepsch2017development, linnenbrink2010measuring}. Surveys were administered using Qualtrics and Brightspace Learning Space. The cognitive load survey consisted of five questions, two of which pertain to the intrinsic cognitive load and three questions pertain to the extraneous cognitive load. Each question in the cognitive load survey was rated on a Likert scale from 1 (Strongly Disagree) to 7 (Strongly Agree). The situational interest survey consisted of three questions assessing context-based interest and attentiveness to the material and was rated on a Likert scale from 1 (Not true at all) to 5 (Very true).

\section{Data Analysis}
Within each of the two experiments, this study defined the instructional modality (Inquiry, Design, and Game) as an independent variable and defined intrinsic cognitive load (ICL), extraneous cognitive load (ECL), and situational interest (SI) as the dependent variables. One-way ANOVA analyses were used to calculate the effect of each of the instructional modalities on each of the dependent variables accounting for variance and statistical significance. It is important to note that both experiments - mechanics and circuits - were designed to be independent of each other. In other words, the two experiments were not intended to be compared with one another; rather, each yielded self-contained results that should be interpreted within its own experimental context.

\section{Results}

\subsection{Experiment 1: Mechanics}
Regarding the first topic in this study - Mechanics - a total of 54 students completed three lab sessions each followed by cognitive load and situational interest surveys. Figure 13 shows the difference in Intrinsic Cognitive Load (ICL) scores across Inquiry, Game, and Design respectively. One-way ANOVA results indicate the there is a statistically significant difference in ICL between the three modalities (p < .001) exceeding a critical value of (F crit = 3.05). These results show that the design modality resulted in the highest ICL with a variance of 1.45, followed by inquiry with a variance of 1.23, then the game which exhibited the lowest ICL with a variance of 1.82. Evidently, these results suggest that the instructional approach significantly influences learners perceived Intrinsic Cognitive Load.

Similar to ICL, the Extraneous Cognitive Load (ECL) was also affected by the type of instructional approach. Figure 14 shows the difference in ECL scores across the three instructional modalities: Inquiry, Game, and Design respectively. ANOVA calculations revealed a statistically significant difference in ECL scores (p = .036) across the three approaches exceeding a critical value of (F crit = 3.05). The design-based lab resulted in the highest level of ECL with a variance of 5.16, followed by the inquiry-based lab with a variance of 5.32, while the game-based lab yielded the lowest ECL with a variance of 6.67.

In addition to ICL and ECL, situational interest was also impacted by the instructional approach. Figure 15 shows the difference in students’ situational interest across the three modalities. ANOVA results revealed a statistically significant effect of instructional approach on situational interest (p = .0096) exceeding a critical value of (F crit = 3.05). The game-based lab showed the highest level of situational interest with a variance of 6.17, followed by the design-based lab with a variance of 5.10, while the inquiry-based lab demonstrated the lowest situational interest scores with a variance of 6.11.

    \begin{figure}[h!]
        \centering
        \includegraphics[width=0.5\textwidth]{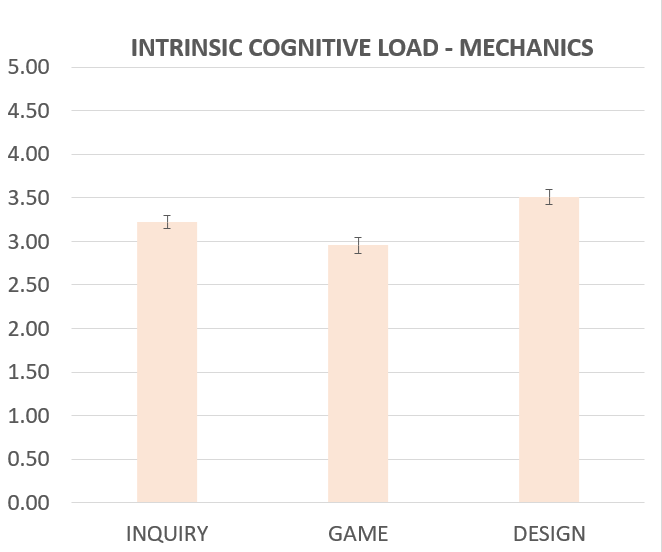}
        \caption{Intrinsic Cognitive Load of Game, Design, and Inquiry in the context of Mechanics}
    \end{figure}

    \begin{figure}[h!]
        \centering
        \includegraphics[width=0.5\textwidth]{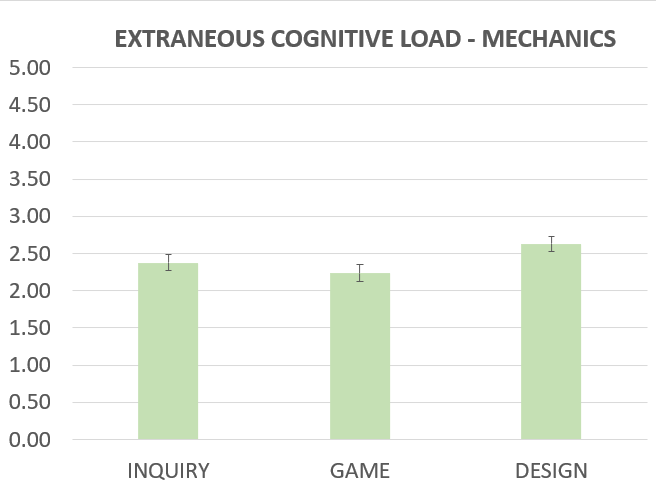}
        \caption{Extraneous Cognitive Load of the 3 lab types in the context of Mechanics}
    \end{figure}

    \begin{figure}[h!]
        \centering
        \includegraphics[width=0.5\textwidth]{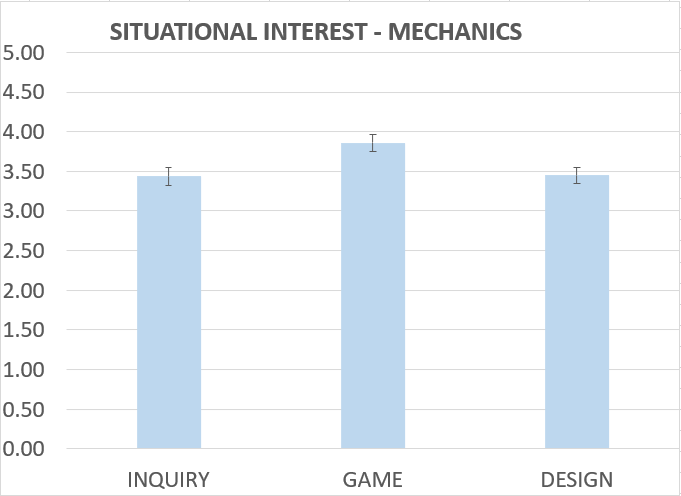}
        \caption{Situational Interest of the 3 lab types in the context of Mechanics}
    \end{figure}
    
\subsection{Experiment 2: Circuits}

In the 2nd experiment, a total of 47 students completed the three labs along with filling out the cognitive load and situational interest surveys after each lab. Figure 16 shows the difference in Intrinsic Cognitive Load (ICL) scores across Inquiry, Game, and Design respectively. One-way ANOVA results indicate that there is a statistically significant difference in ICL between the three lab types (p < .001) which exceeded the critical value (F crit = 3.06). The inquiry-based lab had the highest ICL with a variance of 2.55, followed by the design-based lab with a variance of 1.46. The game-based lab showed the lowest ICL with a variance of 3.11. 

The extraneous cognitive load (ECL) within the topic of Circuits was also significantly affected by the instructional modality as shown in Figure 17. ANOVA analysis revealed a statistically significant difference in ECL (p = .00012) which exceeded the critical value (F crit = 3.06). The inquiry-based lab resulted in the highest ECL with a variance of 8.62, followed by the design-based lab with a variance of 4.77. The game-based lab yielded the lowest ECL with a variance of 7.16.

As expected, the situational interest was also impacted by the lab type within Circuits as shown in Figure 18. ANOVA calculations indicated a statistically significant difference in situational interest (p = .027) exceeding the critical value (F crit = 3.06). The game-based lab produced the highest situational interest scores with a variance of 5.20, followed by the design-based lab with a variance of 5.30. The inquiry-based lab yielded the lowest situational interest outcomes with a variance of 5.42.

    \begin{figure}[h!]
        \centering
        \includegraphics[width=0.5\textwidth]{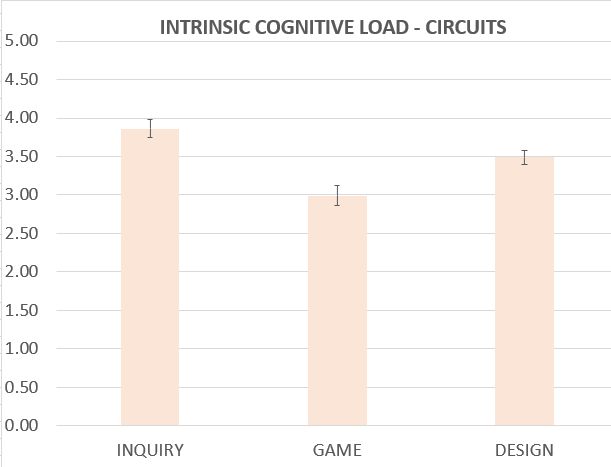}
        \caption{Intrinsic Cognitive Load of the 3 lab types in the context of Circuits}
    \end{figure}

    \begin{figure}[h!]
        \centering
        \includegraphics[width=0.5\textwidth]{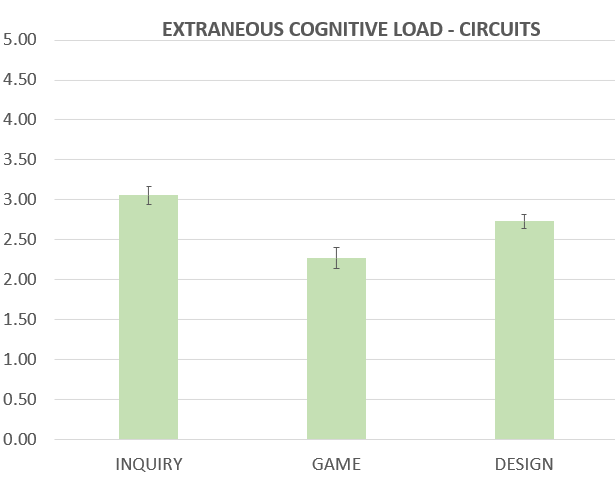}
        \caption{Extraneous Cognitive Load of the 3 lab types in the context of Circuits}
    \end{figure}

    \begin{figure}[h!]
        \centering
        \includegraphics[width=0.5\textwidth]{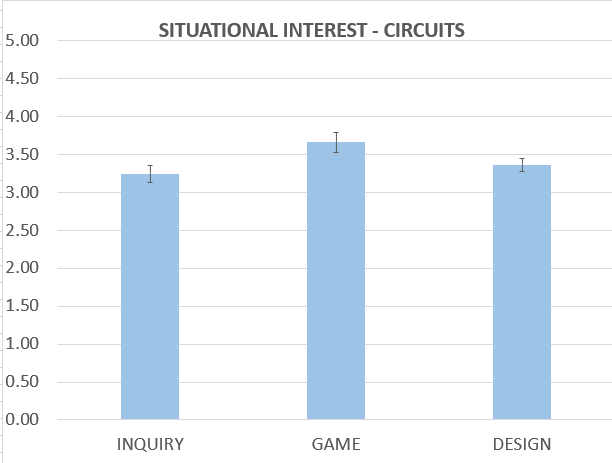}
        \caption{Engagement/ Situational Interest of the 3 lab types in the context of Circuits}
    \end{figure}

\section{Discussion} 

The purpose of this study was to examine how three instructional approaches; design-based, game-based, and inquiry-based learning, differ in their effects on novice learners’ cognitive load (extraneous and intrinsic) and situational interest. Across two experiments, multiple one-way ANOVA analyses were conducted showing a consistent pattern in the results. Below is a breakdown of how the results answer our original research questions and inform practical applications in physics education, along with future steps building on this research.

\subsection{Experiment 1}

In experiment 1, figure 13 shows that the intrinsic cognitive load (ICL) differed significantly across the three instructional approaches. The design-based approach resulted in the highest ICL, followed by the inquiry-based approach, then game-based approach producing the lowest ICL. In the same vein as ICL, the impact of instructional approach on extraneous cognitive load (ECL) was also statistically significant and showed a similar pattern to ICL scores. Figure 14 shows that design-based labs produced the highest levels of ECL and game-based labs produced the lowest ECL while inquiry-based labs landed somewhere in the middle within the context of mechanics.

These findings regarding ICL and ECL indicate that design-based learning, which typically targets problem solving through iterative processes of designing and testing appeared to impose a high level of ICL and ECL than the inquiry-based and game-based approaches in the context of mechanics. This could be attributed to the complexity of design problems since, by nature, they do not have one unique answer or solution \cite{murray2019design} and often involve juggling many variables all at once \cite {roy2008recent}. This is evident especially in mechanics where design usually requires systems thinking where learners must consider how components interact with one another within the whole system \cite{godfrey2014systems}. In that case, learners might also struggle with extra cognitive demands that are irrelevant to the material but rather related to their specific design projects which could hinder their learning as the learners exceed their working memory capacity\cite{sweller2020cognitive, jonassen2010learning}.

Landing second to the design-based approach in ICL and ECL scores was the inquiry-based approach which was perceived as moderately complex and somewhat cognitively demanding which might be due to the self-directed investigation nature of inquiry \cite{van2017inquiry}. Inquiry-based environments often require learners to manage ill-structured problems and integrate multiple sources of information simultaneously \cite{kirschner2006unguided} leading to an increase in intrinsic cognitive load. While these features may support deeper learning, they can also introduce additional processing demands unrelated to the core learning objectives, thereby increasing extraneous cognitive load \cite{surbakti2024cognitive}.

Lastly, the game-based modality appeared to simplify the learning task complexity through structured game play and productive failure, thereby reducing ICL and ECL. This result aligns with the literature on productive failure easing learners' cognitive burden by creating a low-stress environment\cite{kapur2009productive}. Moreover, a study done on the impact of game-based learning on cognitive load showed that game-based instruction is associated with significantly lower ECL and higher learning gains as compared to alternative forms of instruction \cite{schrader2012learning}. Notably, the findings of this experiment align with prior research and builds on it by introducing a systematic comparison between game-based learning and other instructional approaches, namely inquiry-based and design-based learning.

In addition to cognitive load, student scores of situational interest differed significantly across the three instructional approaches within the context of mechanics. Figure 15 shows that the game-based condition produced the highest situational interest, followed by the design-based approach, then the inquiry-based condition yielding the lowest situational interest levels. It is important to notice here that situational interest is associated with low extraneous cognitive load as evident from the game-based lab results. The game-based approach demonstrated low intrinsic and extraneous cognitive load with high situational interest. The situational interest findings agree with prior research that emphasized the role of challenge and enjoyment - which are features of game-based learning — in sustaining learners interest and engagement \cite{hamari2016challenging}.

\subsection{Experiment 2} 

The results of experiment 2 were analogous to those of experiment 1 with slight variances in the effects of design and inquiry on cognitive load. Figure 16 shows that the intrinsic cognitive load (ICL) differed significantly across the three instructional approaches within the context of Circuits. The inquiry-based lab produced the highest ICL, followed by the design-based lab, then the game-based lab producing the lowest ICL. Similar to ICL, the impact of instructional approach on extraneous cognitive load (ECL), was significant. Figure 17 shows that the inquiry-based lab produced the highest ECL, followed by the design-based lab, then the game-based lab resulting in the lowest ECL overall.

The inherent complexities and extra cognitive demands of design and inquiry-based labs channel those in the first experiment. However, inquiry resulted in the highest ICL and ECL in the context of circuits as opposed to design taking the lead in ICL and ECL in the context of mechanics. As established in the literature review section, the domains of mechanics and circuits differ in their respective levels of abstraction and concreteness. Consequently, observed differences in design and inquiry across the two experiments may be attributed to these inherent domain-specific characteristics rather than to the experimental conditions themselves.

Prior research have shown that electrical circuits are characterized as complex tasks because they require the simultaneous integration of multiple interacting elements which can easily increase the intrinsic and extraneous cognitive load \cite{melo2015learning}. The results of this experiment agree with the literature since the inquiry-based circuits lab produced the highest ICL and ECL. Prior studies have also shown that concrete representations in circuits lead to lower intrinsic cognitive load as compared to abstract representations \cite{reisslein2012circuits, melo2015learning}. This work affirms our results which demonstrate that the use of design-based instruction results in lower ICL and ECL than inquiry-based instruction which could be attributed to the nature of circuit design having more concrete and tangible elements than inquiry. 

Showing a very similar pattern to experiment 1, student scores of situational interest differed significantly across the three instructional approaches within the Circuits domain (Figure 18). The game-based condition produced the highest situational interest, followed by the design-based approach, with the inquiry-based condition yielding the lowest situational interest levels. Across both experiments, there was a clear contrast between situational interest and the two types of cognitive load. In Experiment 2, intrinsic and extraneous cognitive load were highest in the inquiry lab, followed by the design lab, and lowest in the game-based lab. In contrast, situational interest followed the opposite pattern, with the game-based lab producing the highest interest, followed by design then inquiry. Taken together, these findings suggest an inverse relationship between cognitive load (both intrinsic and extraneous) and situational interest, such that higher cognitive load tends to be associated with lower situational interest, and vice versa.

\subsection{Relationship between Cognitive Load and Situational Interest}

As seen from ICL, ECL, and SI graphs, there appears to be a correlation between low cognitive load and high situational interest. This was especially evident and significant in the game-based lab in both experiments since the game modality consistently yielded the lowest cognitive load (intrinsic and extraneous) and the highest situational interest. The relationship between cognitive load and situational interest may be explained by the fact that lower levels of extraneous and intrinsic cognitive load free up cognitive resources, enabling learners to focus more fully on the learning activity and, in turn, experience greater involvement and persistence leading to higher interest levels \cite{evans2024cognitive}. Conversely, higher intrinsic and extraneous cognitive load may lead to lower engagement levels as seen in the inquiry and design-based labs which produced high cognitive load with comparatively lower interest than the game-based labs. This suggests that the cognitive demands of inquiry and design may have exceeded some learners’ optimal challenge levels, thereby reducing engagement. 

Given the results of this study, it is important to clarify that the findings do not imply that one instructional modality is inherently better than the others. Instead, the results point to the role of the inherent complexity of each modality and the additional elements they introduce to the learning environment. These features should help guide instructors in making informed decisions about how and when to integrate different modalities in physics instruction. Considering the fit between a given modality, the instructional goals, and students’ cognitive demands may support more meaningful engagement and deeper conceptual understanding. Furthermore, it is also important to note that the specific design of the educational game influences the extent to which learners experience higher or lower cognitive demands. Well-designed educational games often embed instructional content within intuitive rules and clear goal structures, which may reduce unnecessary cognitive processing \cite{habgood2011motivating}.

\section{limitations and future work}

One limitation of this study concerns the sequencing of laboratory activities within each experiment. Although three distinct types of laboratory sessions were implemented, their order was not systematically controlled in both experiments. As a result, sequencing effects may have influenced learners’ engagement and cognitive load, thereby complicating the interpretation of the findings. Furthermore, the study did not account for participants’ prior knowledge, which may have interacted with both cognitive load and situational interest. Future research will address these limitations by systematically varying the order of laboratory activities and examining the role of students' prior knowledge across multiple physics domains.

\section{Conclusion}

This study highlights the importance of evaluating physics laboratory instruction for non-STEM majors through an integrated cognitive and motivational perspective. By drawing on cognitive load theory and the situational interest framework, this research offers insight into how the instructional modality influences not only the cognitive demands that novice physics learners experience but also the degree to which they find physics laboratories engaging and meaningful. By comparing design-based, inquiry-based, and game-based approaches in contexts of mechanics and circuits this study systematically explores how students’ cognitive effort and their interest in the physics lab influence one another across different instructional settings. This work seeks to inform the development of instructional approaches that enhance the quality of physics learning experiences for novice students.

Overall, this study contributes to the growing body of research advocating for diversified instructional strategies that consider both cognitive and affective dimensions of learning. It also adds to the literature on teaching and learning different physics domains like mechanics and circuits based on the level of concreteness and abstractness of each domain. Through understanding the context and the required learning resources each instructional modality offers, we can foster a more meaningful and engaging learning experience in undergraduate physics education for learners coming from non-STEM fields. Such an approach acknowledges that novice learners may benefit from varying instructional methods depending on the topic and context. Ultimately, this study aims to offer a hopeful path towards more supportive, engaging, and meaningful physics instruction. 

\section{Acknowledgments}
ChatGPT-4.0 was employed for improving word choices and sentence structure as well as grammar checks. It was not used for any ideation or data analysis.
\clearpage
\bibliography{references}

@book{kalyuga2025rethinking,
  title={Rethinking Cognitive Load Theory},
  author={Kalyuga, Slava and others},
  year={2025},
  publisher={Oxford University Press}
}

@article{kirschner2006unguided,
  title={Why unguided learning does not work: An analysis of the failure of discovery learning, problem-based learning, experiential learning and inquiry-based learning},
  author={Kirschner, Paul and Sweller, John and Clark, Richard E},
  journal={Educational psychologist},
  volume={41},
  number={2},
  pages={75--86},
  year={2006}
}

@article{sweller1988cognitive,
  title={Cognitive load during problem solving: Effects on learning},
  author={Sweller, John},
  journal={Cognitive science},
  volume={12},
  number={2},
  pages={257--285},
  year={1988},
  publisher={Elsevier}
}

@article{sweller2003evolution,
  title={Evolution of human cognitive architecture},
  author={Sweller, John},
  journal={Psychology of learning and motivation},
  volume={43},
  pages={216--266},
  year={2003}
}

@incollection{sweller2011cognitive,
  title={Cognitive load theory},
  author={Sweller, John},
  booktitle={Psychology of learning and motivation},
  volume={55},
  pages={37--76},
  year={2011},
  publisher={Elsevier}
}

@article{klepsch2017development,
  author = {Klepsch, M. and Schmitz, F. and Seufert, T.},
  title = {Development and validation of two instruments measuring intrinsic, extraneous, and germane cognitive load},
  journal = {Frontiers in Psychology},
  volume = {8},
  pages = {1997},
  year = {2017}
}

@book{kapp2012gamification,
  title={The gamification of learning and instruction: game-based methods and strategies for training and education},
  author={Kapp, Karl M},
  year={2012},
  publisher={John Wiley \& Sons}
}

@article{linnenbrink2010measuring,
  title={Measuring situational interest in academic domains},
  author={Linnenbrink-Garcia, Lisa and Durik, Amanda M and Conley, AnneMarie M and Barron, Kenneth E and Tauer, John M and Karabenick, Stuart A and Harackiewicz, Judith M},
  journal={Educational and psychological measurement},
  volume={70},
  number={4},
  pages={647--671},
  year={2010},
  publisher={Sage Publications Sage CA: Los Angeles, CA}
}

@book{jonassen2010learning,
  title={Learning to solve problems: A handbook for designing problem-solving learning environments},
  author={Jonassen, David H},
  year={2010},
  publisher={Routledge}
}

@article{chang2017development,
  title={Development of an effective educational computer game based on a mission synchronization-based peer-assistance approach},
  author={Chang, Shao-Chen and Hwang, Gwo-Jen},
  journal={Interactive Learning Environments},
  volume={25},
  number={5},
  pages={667--681},
  year={2017},
  publisher={Taylor \& Francis}
}

@article{perttula2017flow,
  title={Flow experience in game based learning--a systematic literature review},
  author={Perttula, Arttu and Kiili, Kristian and Lindstedt, Antero and Tuomi, Pauliina},
  journal={International Journal of Serious Games},
  volume={4},
  number={1},
  pages={57--72},
  year={2017}
}

@article{sweller2022role,
  title={The role of evolutionary psychology in our understanding of human cognition: Consequences for cognitive load theory and instructional procedures},
  author={Sweller, John},
  journal={Educational Psychology Review},
  volume={34},
  number={4},
  pages={2229--2241},
  year={2022},
  publisher={Springer}
}

@article{kalyuga2011cognitive,
  title={Cognitive load theory: How many types of load does it really need?},
  author={Kalyuga, Slava},
  journal={Educational psychology review},
  volume={23},
  number={1},
  pages={1--19},
  year={2011},
  publisher={Springer}
}

@article{debue2014does,
  title={What does germane load mean? An empirical contribution to the cognitive load theory},
  author={Debue, Nicolas and Van De Leemput, C{\'e}cile},
  journal={Frontiers in psychology},
  volume={5},
  pages={1099},
  year={2014},
  publisher={Frontiers Media SA}
}

@article{homer2014moved,
  title={Moved to learn: The effects of interactivity in a Kinect-based literacy game for beginning readers},
  author={Homer, Bruce D and Kinzer, Charles K and Plass, Jan L and Letourneau, Susan M and Hoffman, Dan and Bromley, Meagan and Hayward, Elizabeth O and Turkay, Selen and Kornak, Yolanta},
  journal={Computers \& Education},
  volume={74},
  pages={37--49},
  year={2014},
  publisher={Elsevier}
}

@article{kapur2009productive,
  title={Productive failure in CSCL groups},
  author={Kapur, Manu and Kinzer, Charles K},
  journal={International Journal of Computer-Supported Collaborative Learning},
  volume={4},
  number={1},
  pages={21--46},
  year={2009},
  publisher={Springer}
}

@article{plass2015foundations,
  title={Foundations of game-based learning},
  author={Plass, Jan L and Homer, Bruce D and Kinzer, Charles K},
  journal={Educational psychologist},
  volume={50},
  number={4},
  pages={258--283},
  year={2015},
  publisher={Taylor \& Francis}
}

@book{plass2020handbook,
  title={Handbook of game-based learning},
  author={Plass, Jan L and Mayer, Richard E and Homer, Bruce D},
  year={2020},
  publisher={Mit Press}
}

@article{plass2019four,
  title={Four ways of considering emotion in cognitive load theory},
  author={Plass, Jan L and Kalyuga, Slava},
  journal={Educational psychology review},
  volume={31},
  number={2},
  pages={339--359},
  year={2019},
  publisher={Springer}
}

@article{hawthorne2025relationship,
  title={The relationship between positive and painful emotions and cognitive load during an algebra learning task},
  author={Hawthorne, Benjamin S and Slemp, Gavin R and Vella-Brodrick, Dianne A and Hattie, John},
  journal={Learning and Individual Differences},
  volume={117},
  pages={102597},
  year={2025},
  publisher={Elsevier}
}

@book{hidi2004interest,
  title={Interest, a motivational variable that combines affective and cognitive functioning},
  author={Hidi, Suzanne and Renninger, K and Krapp, Andreas},
  year={2004},
  publisher={Lawrence Erlbaum}
}

@article{renninger2011revisiting,
  title={Revisiting the conceptualization, measurement, and generation of interest},
  author={Renninger, K Ann and Hidi, Suzanne},
  journal={Educational psychologist},
  volume={46},
  number={3},
  pages={168--184},
  year={2011},
  publisher={Taylor \& Francis}
}

@article{rotgans2011situational,
  title={Situational interest and academic achievement in the active-learning classroom},
  author={Rotgans, Jerome I and Schmidt, Henk G},
  journal={Learning and instruction},
  volume={21},
  number={1},
  pages={58--67},
  year={2011},
  publisher={Elsevier}
}

@article{schiefele1991interest,
  title={Interest, learning, and motivation},
  author={Schiefele, Ulrich},
  journal={Educational psychologist},
  volume={26},
  number={3-4},
  pages={299--323},
  year={1991},
  publisher={Taylor \& Francis}
}

@online{energysimulation,
  author    = {Phet},
  title     = {Energy Simulation},
  url       = {https://phet.colorado.edu/sims/html/energy-skate-park-basics/latest/energy-skate-park-basics_en.html},
  note      = {Accessed: 15 January 2026}
}

@online{circuitssimulation,
  author    = {Phet},
  title     = {Circuits Simulation},
  url       = {https://phet.colorado.edu/sims/html/circuit-construction-kit-dc/latest/circuit-construction-kit-dc_en.html},
  note      = {Accessed: 15 January 2026}
}

@online{rollercoaster,
  author    = {Compassproject},
  title     = {Roller Coaster},
  url       = {https://sims.compassproject.net/rollercoaster/rollercoaster_en.html},
  note      = {Accessed: 15 January 2026}
}

@book{kalmpourtzis2018educational,
  title={Educational Game Design Fundamentals: A journey to creating intrinsically motivating learning experiences},
  author={Kalmpourtzis, George},
  year={2018},
  publisher={AK Peters/CRC Press}
}

@article{coletta2019feeling,
  title={Feeling Newton’s second law},
  author={Coletta, Vincent P and Bernardin, Josh and Pascoe, Daniel and Hoemke, Anatol},
  journal={The Physics Teacher},
  volume={57},
  number={2},
  pages={88--90},
  year={2019},
  publisher={AIP Publishing}
}

@article{dale2020step,
  title={A step forward in kinesthetic activities for teaching kinematics in introductory physics},
  author={Dale, Zachary and DeStefano, Paul R and Shaaban, Lori and Siebert, Cora and Widenhorn, Ralf},
  journal={American Journal of Physics},
  volume={88},
  number={10},
  pages={825--830},
  year={2020},
  publisher={AIP Publishing}
}

@article{vieyra2024making,
  title={Making Motion Meaningful: Mapping Body Movements onto Graphs},
  author={Vieyra, Rebecca E and Megowan Romanowicz, Colleen and Johnson-Glenberg, Mina C and O’Brien, Daniel and Vieyra Cort{\'e}s, Chrystian},
  journal={The Science Teacher},
  volume={91},
  number={6},
  pages={57--64},
  year={2024},
  publisher={Taylor \& Francis}
}

@article{solomon2022embodied,
  title={Embodied physics: Utilizing dance resources for learning and engagement in STEM},
  author={Solomon, Folashad{\'e} and Champion, Dionne and Steele, Mariah and Wright, Tracey},
  journal={Journal of the Learning Sciences},
  volume={31},
  number={1},
  pages={73--106},
  year={2022},
  publisher={Taylor \& Francis}
}

@article{bauman2024student,
  title={Student conceptual resources for understanding electric circuits},
  author={Bauman, Lauren C and Hansen, Brynna and Goodhew, Lisa M and Robertson, Amy D},
  journal={Physical Review Physics Education Research},
  volume={20},
  number={2},
  pages={020128},
  year={2024},
  publisher={APS}
}

@inproceedings{hazelton2013assessing,
  title={Assessing the impact of a computer simulation in conjunction with tutorials in introductory physics on conceptual understanding},
  author={Hazelton, RL and Shaffer, Peter S and Heron, PR},
  booktitle={Proceedings of the Physics Education Research Conference},
  pages={177--180},
  year={2013}
}

@article{tomkelski2023physics,
  title={Physics Teachers' Learning on the Use of Multiple Representations in Lesson Study about Ohm's Law.},
  author={Tomkelski, Mauri Lu{\'\i}s and Baptista, M{\'o}nica and Richit, Adriana},
  journal={European Journal of Science and Mathematics Education},
  volume={11},
  number={3},
  pages={427--444},
  year={2023},
  publisher={ERIC}
}

@article{robertson2023identifying,
  title={Identifying student conceptual resources for understanding physics: A practical guide for researchers},
  author={Robertson, Amy D and Goodhew, Lisa M and Bauman, Lauren C and Hansen, Brynna and Alesandrini, Anne T},
  journal={Physical Review Physics Education Research},
  volume={19},
  number={2},
  pages={020138},
  year={2023},
  publisher={APS}
}

@article{chiu2005promoting,
  title={Promoting fourth graders' conceptual change of their understanding of electric current via multiple analogies},
  author={Chiu, Mei-Hung and Lin, Jing-Wen},
  journal={Journal of Research in Science Teaching: The Official Journal of the National Association for Research in Science Teaching},
  volume={42},
  number={4},
  pages={429--464},
  year={2005},
  publisher={Wiley Online Library}
}

@article{roy2008recent,
  title={Recent advances in engineering design optimisation: Challenges and future trends},
  author={Roy, Rajkumar and Hinduja, Srichand and Teti, Roberto},
  journal={CIRP annals},
  volume={57},
  number={2},
  pages={697--715},
  year={2008},
  publisher={Elsevier}
}

@article{murray2019design,
  title={Design by taking perspectives: How engineers explore problems},
  author={Murray, Jaclyn K and Studer, Jaryn A and Daly, Shanna R and McKilligan, Seda and Seifert, Colleen M},
  journal={Journal of Engineering Education},
  volume={108},
  number={2},
  pages={248--275},
  year={2019},
  publisher={Wiley Online Library}
}

@article{van2017inquiry,
  title={Inquiry-based science education: Scaffolding pupils’ self-directed learning in open inquiry},
  author={Van Uum, Martina SJ and Verhoeff, Roald P and Peeters, Marieke},
  journal={International Journal of Science Education},
  volume={39},
  number={18},
  pages={2461--2481},
  year={2017},
  publisher={Taylor \& Francis}
}

@article{godfrey2014systems,
  title={Systems thinking, systems design and learning power in engineering education},
  author={Godfrey, Patrick and Crick, R Deakin and Huang, Shaofu},
  journal={International Journal of Engineering Education},
  year={2014}
}

@article{sweller2020cognitive,
  title={Cognitive load theory and educational technology},
  author={Sweller, John},
  journal={Educational technology research and development},
  volume={68},
  number={1},
  pages={1--16},
  year={2020},
  publisher={Springer}
}

@article{surbakti2024cognitive,
  title={Cognitive load theory: Implications for instructional design in digital classrooms},
  author={Surbakti, Rudy and Umboh, Satria Evans and Pong, Ming and Dara, Sokha},
  journal={International Journal of Educational Narratives},
  volume={2},
  number={6},
  pages={483--493},
  year={2024}
}

@article{schrader2012learning,
  title={Learning in educational computer games for novices: The impact of support provision types on virtual presence, cognitive load, and learning outcomes},
  author={Schrader, Claudia and Bastiaens, Theo},
  journal={International Review of Research in Open and Distributed Learning},
  volume={13},
  number={3},
  pages={206--227},
  year={2012},
  publisher={{\'E}rudit}
}

@article{hamari2016challenging,
  title={Challenging games help students learn: An empirical study on engagement, flow and immersion in game-based learning},
  author={Hamari, Juho and Shernoff, David J and Rowe, Elizabeth and Coller, Brianno and Asbell-Clarke, Jodi and Edwards, Teon},
  journal={Computers in human behavior},
  volume={54},
  pages={170--179},
  year={2016},
  publisher={Elsevier}
}

@article{evans2024cognitive,
  title={Cognitive load theory and its relationships with motivation: A self-determination theory perspective},
  author={Evans, Paul and Vansteenkiste, Maarten and Parker, Philip and Kingsford-Smith, Andrew and Zhou, Sijing},
  journal={Educational Psychology Review},
  volume={36},
  number={1},
  pages={7},
  year={2024},
  publisher={Springer}
}

@article{reisslein2012circuits,
  title={Circuits kit K--12 outreach: Impact of circuit element representation and student gender},
  author={Reisslein, Jana and Ozogul, Gamze and Johnson, Amy M and Bishop, Kristen L and Harvey, Justin and Reisslein, Martin},
  journal={IEEE Transactions on Education},
  volume={56},
  number={3},
  pages={316--321},
  year={2012},
  publisher={IEEE}
}

@article{melo2015learning,
  title={Learning electrical circuits: the effects of the 4C-ID instructional approach in the acquisition and transfer of knowledge.},
  author={Melo, Mario and Miranda, Guilhermina Lobato},
  journal={Journal of Information Technology Education: Research},
  volume={14},
  year={2015}
}

@article{habgood2011motivating,
  title={Motivating children to learn effectively: Exploring the value of intrinsic integration in educational games},
  author={Habgood, MP Jacob and Ainsworth, Shaaron E},
  journal={The Journal of the Learning Sciences},
  volume={20},
  number={2},
  pages={169--206},
  year={2011},
  publisher={Taylor \& Francis}
}

\end{document}